\documentclass[twoside]{JHEP3}
\pdfoutput=1
\usepackage{epsfig,multicol,slashed}
\usepackage{delarray,amsmath,bbm}

\usepackage{amssymb}
\usepackage{amsthm}
\usepackage{dcolumn}
\usepackage{graphics}
\usepackage{graphicx}
\usepackage{longtable}

\newcommand{\nn}{\nonumber}
\newcommand{\beq} {\begin{equation}}
\newcommand{\eeq} {\end{equation}}
\newcommand{\beqa} {\begin{eqnarray}}
\newcommand{\eeqa} {\end{eqnarray}}

\newcommand{\ie}{{\it i.e.}}
\newcommand{\eg}{{\it e.g.}}

\newcommand{\wrt}{{\it wrt.\ }}
\newcommand{\rhs}{{\it rhs.\ }}

\newcommand{\as}{{\alpha_s}}
\newcommand{\lqcd}{\Lambda_{QCD}}
\newcommand{\chat}{\hat{\bs{\ell}}}
\newcommand{\la}{\Lambda}

\newcommand{\ieps}{i\varepsilon}

\newcommand{\order}[1]{${\cal O}\left(#1 \right)$}
\newcommand{\morder}[1]{{\cal O}\left(#1 \right)}
\newcommand{\eq}[1]{(\ref{#1})}
\newcommand{\fig}[1]{Fig.~\ref{#1}}

\newcommand{\inv}[1]{\frac{1}{#1}}
\newcommand{\ket}[1]{\vert{#1}\rangle}
\newcommand{\bra}[1]{\langle{#1}\vert}

\newcommand{\com}[2]{\left[{#1},{#2}\right]}
\newcommand{\acom}[2]{\left\{{#1},{#2}\right\}}
\newcommand{\tr}{\mathrm{Tr}\,}

\newcommand{\bs}[1]{\boldsymbol{#1}}

\newcommand{\mA}{\mathcal{A}}

\newcommand{\mD}{\mathcal{D}}

\newcommand{\xv}{{\bs{x}}}
\newcommand{\yv}{{\bs{y}}}
\newcommand{\pv}{{\bs{p}}}
\newcommand{\kv}{{\bs{k}}}
\newcommand{\qv}{{\bs{q}}}

\newcommand{\Av}{{\bs{A}}}
\newcommand{\Yv}{{\bs{Y}}}
\newcommand{\gv}{\bs{\gamma}}

\newcommand{\nabv}{{\bs{\nabla}}}

\newcommand{\qu}{{\rm q}}
\newcommand{\qb}{{\rm\bar q}}

\newcommand{\halft}{{\textstyle \frac{1}{2}}}
\newcommand{\thirt}{{\textstyle \frac{1}{3}}}
\newcommand{\quart}{{\textstyle \frac{1}{4}}}
\newcommand{\lsim}{\lesssim}
\newcommand{\gsim}{\gtrsim}

\title{\center{Introduction to QCD\thanks{Based on lectures at the International Summer School and Conference on High Energy Physics: Standard Model and Beyond (ISSCSMB '10), at Mugla, Akyaka in Turkey on 27 August -- 4 September 2010.}\\ {\rm\Large -- a bound state perspective\ }}}
\author{Paul Hoyer\\
              Department of Physics and Helsinki Institute of
              Physics\\
              POB 64, FIN-00014 University of Helsinki, Finland \\
              }

\abstract{These lecture notes focus on the bound state sector of QCD. Motivated by data which suggests that the strong coupling $\as(Q)$ freezes at low $Q$, and by similarities between the spectra of hadrons and atoms, I discuss if and how QCD bound states may be treated perturbatively. I recall the basic principles of perturbative gauge theory bound states at lowest order in the $\hbar$ expansion. Born level amplitudes are insensitive to the $\ieps$ prescription of propagators, which allows to eliminate the $Z$-diagrams of relativistic, time-ordered Coulomb interactions. The Dirac wave function thus describes a single electron which propagates forward in time only, even though the bound state has any number of pair constituents when Feynman propagators are used. In the absence of an external potential, states that are bound by the Coulomb attraction of their constituents can be analogously described using only their valence degrees of freedom. The instantaneous  $A^0$ field is determined by Gauss' law for each wave function component, \ie, for each position of the valence constituents. Solutions for $A^0$ obtained with a boundary condition that imposes an asymptotically constant energy density give rise to a linear potential for color singlet $\qu\qb$ and $\qu\qu\qu$ states. The strength of the linear potential is determined by the boundary condition and is of lower order in $\as$ than the gluon exchange interaction, which may then be treated as a higher order perturbative correction. Bound states evaluated to a given order in $\as$ and $\hbar$ must have the full symmetry of the exact theory, including the dynamic boost invariance of states quantized at equal time. The wave functions are indeed found to have such a hidden invariance, which ensures the correct dependence of the energy eigenvalues on the center of mass momentum. Thus relativistic bound states can be studied using perturbative methods.}

\begin{document}

\section{The strong interactions}

There is a broad consensus that Quantum Chromodynamics is {\it the} theory underlying the strong interactions. This conviction first arose from general features of QCD which agree with observations. The realization that the QCD coupling $\as$ decreases with momentum transfer (asymptotic freedom) was decisive for establishing QCD as a serious candidate theory in 1973. QCD was also seen to have approximate isospin and chiral symmetry for small quark masses. The theory later has had an impressive success in describing a large variety of data on short-distance processes. These involve small $\as$ and thus directly probe the QCD lagrangian via the perturbative expansion. Numerical methods based on discrete lattice approximations of space-time have demonstrated that the properties of QCD at long distances, in particular its bound states (the hadrons), are at least qualitatively in agreement with observations.

What I summarize above in a few sentences would provide material for many lecture courses. The vast amount of information has fortunately inspired a ``Resource Letter'' \cite{Kronfeld:2010bx}, which at 39 pages is no letter but a comprehensive collection of references to the QCD literature. It covers all aspects of QCD and guides to material organized by scope and depth via brief remarks. Many excellent introductory lectures on QCD have been published as well, see for example \cite{Hollik:2010id,Pich:2007vu,Seymour:2005hs}. I shall not try to rival these here. Instead, I have chosen to focus on a less well covered but fascinating aspect: the bound states.

Color confinement and the spontaneous breaking of chiral symmetry are novel features of QCD, whose origin is still poorly understood. Both are closely connected to the physics of hadrons as bound states of QCD. Color confinement implies that only color neutral states (hadrons and nuclei) propagate macroscopic distances and thus reach experimental detectors. The spontaneous breaking of chiral symmetry explains the absence of parity doubling in the hadron spectrum and gives a reason for the small mass of the pion. Hadrons also have other puzzling properties: {\em Hard scattering} shows that they have a large (infinite) number of highly relativistic quark and gluon constituents, whereas the {\em hadron spectrum} reflects just the valence quark ($\qu\qb$ or $\qu\qu\qu$) degrees of freedom \cite{Nakamura:2010zzi}. This property makes hadrons quite different from more familiar bound states, such as atoms and molecules, whose spectra reflect the relative motion (vibrations, rotations,...) of all their constituents.

The $u,d$ valence quark masses contribute $\lsim 2\%$ to the nucleon mass. Hadrons are in fact the only truly relativistic bound states found in Nature. In QED atoms the relativistic effects are of \order{\alpha} and can be calculated perturbatively with exquisite precision \cite{Jentschura:2005}. A relativistic description of hadrons would seem to require $\as\gsim 1$ and thus methods beyond perturbation theory. Together with the increase of the running coupling $\as(Q)$ at low momentum scales $Q$ (``infrared slavery'') this has led to the general view that QCD is strongly coupled at distances $\sim 1$ fm.

But data suggests otherwise. I review in Sec. \ref{freeze} the indications that $\as(Q)$ freezes at a modest value $\alpha_0\sim 0.5$ as $Q\to 0$ and that QCD remains perturbative at long distance. There is little freedom in a perturbative approach since the lagrangian determines the expansion nearly uniquely, and qualitatively correct results should be obtained already at lowest order in $\as$. Hence the relevance of perturbation theory for soft QCD physics is an assertion that can be addressed -- and potentially excluded. Nevertheless, aspects of relativistic bound states that definitely {\it can} be addressed perturbatively receive little attention:
\begin{itemize}
\item[-] Does the Hydrogen atom wave function Lorentz contract? \cite{Jarvinen:2004pi}
\item[-] How does the Dirac wave function describe higher ($e^+e^-$ pair) Fock states? \cite{Hoyer:2009ep}
\item[-] Is there a Born term for bound states, as there is for scattering amplitudes? \cite{Brodsky:2010zk}
\end{itemize}
Modern courses in field theory devote surprisingly little attention to bound states. The perturbative description of bound states differs qualitatively from that of scattering amplitudes, since bound state poles arise through the {\it divergence} of the perturbative expansion. It thus seems worthwhile to devote these lectures to bound state issues. I shall not attempt an objective overview, but rather focus on issues which seem relevant for a possible description of QCD bound states.

I discuss the evidence that QCD remains perturbative at long distances in Sec. \ref{freeze}. Sec.~\ref{boundpt} introduces general issues related to the perturbative description of bound states. The possibility of using retarded propagators at Born level and how this affects wave functions is discussed in Sec. \ref{retprop}, and then applied to Dirac bound states in Sec. \ref{dirwf}. Sec. \ref{dynboost} addresses the dynamical (as opposed to manifest) boost invariance of states defined at equal time of their constituents. Each order in an $\hbar$ expansion must be Lorentz invariant, as discussed in Sec. \ref{hbarsect}. In gauge theories the $A^0$ potential is at each instant of time determined by the charges through Gauss' law. In Sec. \ref{gausssect} I note that the Coulomb field of the Hydrogen atom therefore is different for each wave function component (position of the charges). This allows me in Sec. \ref{confbound} to consider solving Gauss' law with a boundary condition where $A^0$ does not vanish at spatial infinity. This generates a linear confining potential without breaking the symmetries of the theory. In Secs. \ref{mesons} and \ref{baryons} I use this boundary condition to derive the QCD bound state equations at lowest order in $\hbar$ and $\as$ for $\qu\qb$ and $\qu\qu\qu$ states. Final comments are given in Sec. \ref{discuss}.

\section{The freezing of $\as(Q)$ \label{freeze}}

Strong interaction phenomena are characterized by the scale $\lqcd \simeq 200\ {\rm MeV} \simeq 1\ {\rm fm}^{-1}$. This fundamental constant also determines the value of $\as$ in the perturbative regime: $\as(M_Z)=0.1184 \pm 0.0007$ \cite{Bethke:2009jm}. The running of $\as(Q)$ shown in \fig{fig1} (left panel) agrees well with data down to the scale of the $\tau$ lepton mass, where $\as(1.8\ {\rm GeV}) \simeq 0.33$. 

%
\EPSFIGURE[ht]{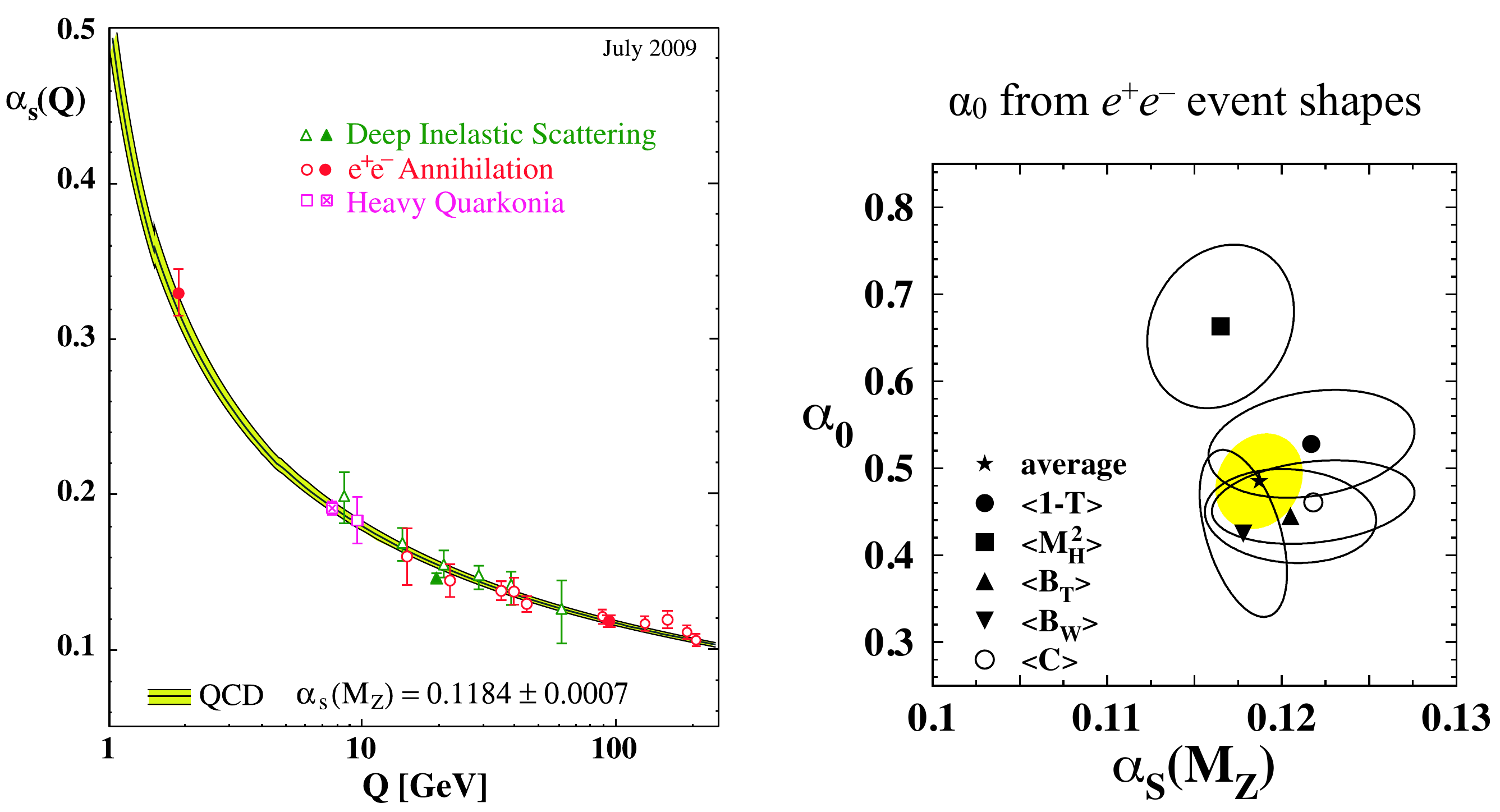,width=1.\columnwidth}{{\em Left:} The dependence of the QCD coupling $\as$ on the momentum scale $Q$ \cite{Bethke:2009jm}. {\em Right:} The average coupling $\alpha_0$ defined by \eq{alpha0} as determined from data in the range $0 < Q < 2$ GeV \cite{MovillaFernandez:2001ed}.
\label{fig1}}
%

It is often claimed that $\as(Q)$ grows large as $Q \to \lqcd$, and that confinement is a consequence of strongly coupled QCD. Actually, several analyses indicate that $\as(Q)$ ``freezes'', \ie, becomes independent of $Q$ at low scales \cite{Brodsky:2002nb}. A dispersive approach \cite{Dokshitzer:1995qm} which uses moments of event shapes to extract an average coupling $\alpha_0$ at low scales,
\beq\label{alpha0}
\alpha_0(\mu_I) \equiv \inv{\mu_I}\int_0^{\mu_I}dQ\, \as(Q)
\eeq
gave \cite{MovillaFernandez:2001ed} $\alpha_0 \simeq 0.5$ (for $\mu_I=2$ GeV) as shown in the right panel of \fig{fig1}. A recent analysis of event shapes in $e^+e^-$ annihilations which combined the dispersive method with perturbative calculations at NNLO accuracy resulted in \cite{Gehrmann:2009eh}
\beqa
\as(M_Z) &=& 0.1153 \pm 0.0017\;({\rm exp}) \pm 0.0023\;({\rm th})\nn\\
\alpha_0 &=& 0.5132 \pm 0.0115\;({\rm exp}) \pm 0.0381\;({\rm th}) 
\eeqa

A moderate coupling at long distances would help explain why perturbative results are found to qualitatively describe data at low $Q^2$. Examples include Bloom-Gilman duality \cite{Malace:2009dg}, precocious dimensional scaling \cite{Anderson:1976ph} and the inclusive distribution of hadrons in and between jets \cite{Fong:1990nt}. Hadrons are, like atoms, classified according to the spin and orbital angular momentum of their valence constituents. The success of the quark model in describing hadron masses, magnetic moments and other properties suggests a perturbative structure. In the words of a well-known expert \cite{Dokshitzer:1998nz}: 
\begin{quote}
\begin{center}
QCD is about to undergo a {\it faith transition}.
\end{center}
 QCD practitioners prepare themselves -- slowly but steadily -- to 
start using, in earnest, the language of {\it quarks} and {\it gluons} down 
into the region of small characteristic momenta -- {\it ``large distances''.}
\end{quote}
This is a perplexing statement since we \emph{know} that QCD is not perturbative in the same sense as QED is -- there are no free quarks or gluons. Applied to the phenomenological quark model potential 
\beq\label{potential}
V(r) = kr-\frac{4}{3}\frac{\as}{r}
\eeq
it suggests that the coefficient $k$ of the linear term is of lower order in $\as$ than the gluon exchange contribution $-\frac{4}{3}\as/r$. In Sec. \ref{confbound} I show  that a linear term arises in the perturbative expansion if a non-vanishing energy density $\propto \lqcd^4$ is imposed on the solution of Gauss' law. Then $k \propto \sqrt{\as}\,\lqcd^2$ is indeed of lower order than the \order{\as} gluon exchange contribution, which may be treated as a higher order perturbative correction.

\section{Bound states in perturbation theory \label{boundpt}}

Bound states appear as poles in scattering amplitudes. In a perturbative expansion the poles are generated by the {\em divergence} of the sum -- no finite order Feynman diagram has a bound state singularity. This is intuitively understandable, since the constituents must continuously interact in order to stay bound. It may nevertheless seem surprising that the QED perturbative series diverges even for atoms, however small is $\alpha$ (I return to this in Sec. \ref{hbarsect}). Having to sum an infinite set of diagrams raises the question of which diagrams to include in the sum: Unlike in ordinary perturbation theory one cannot simply order the calculation according to the power of the coupling $\alpha$. Summing different sets of diagrams leads to different approximations for the bound states,  which {\em a priori} appear equally justified \cite{Lepage:1978hz}.

The perturbative expansion of scattering amplitudes can be viewed not only as a power series in $\alpha$, but also as a loop expansion. Loops are associated with powers of the Planck constant $\hbar$  \cite{Itzykson:1980rh}. Born terms may have different powers of $\alpha$ depending on the process, but are always of lowest order in $\hbar$ in the sense that they have no loops. The very successful QED calculations of non-relativistic atoms are also based on a loop expansion. The loops are counted in the interaction kernel, which is then iterated as a geometric series to generate the bound states. Does this mean that the Schr\"odinger equation, which follows from iterating single photon exchange with no loop correction, is the Born approximation for non-relativistic bound states? And if so, can this concept of a Born term be extended to relativistic bound states, defining a specific lowest order approximation (bound state equation)? The $\hbar$ expansion has received rather little attention in a field theory context, but it appears that the answer is positive to both questions. I discuss the $\hbar$ expansion further in Sec. \ref{hbarsect}.

Having identified the Born term one may ask under what circumstances this approximation to a bound state is reliable. Unsurprisingly, it turns out that this requires the coupling $\alpha$ to be small enough for a perturbative (loop) expansion of the kernel to be well motivated. At first this appears to make the Born term concept useless for relativistic states. In QED atoms the constituents move with speeds $v/c \simeq \alpha$. Hence relativistic motion requires a large coupling $\alpha \gsim 1$, which invalidates the loop expansion. The Dirac equation with a $-\alpha/r$ Coulomb potential in fact gives complex bound state energies for $\alpha > 1$ \cite{Itzykson:1980rh}.

A systematic study of bound states using perturbation theory requires expanding both in $\hbar$ (which at lowest order defines the Born term) and in the coupling $\alpha$ (or $\as$ in QCD). The general understanding (which I review in Sec. \ref{hbarsect}) is that these two expansions are coupled: Each loop adds a power of $\hbar$ and a power of $\alpha$. At small $\alpha$ we then have non-relativistic states described by the Schr\"odinger equation -- unless there is a confining potential at lowest order in $\alpha$. In Sec. \ref{confbound} I argue that the perturbative expansion does allow a linear potential, through a non-vanishing boundary condition in Gauss' law.

In gauge theories the $A^0$ field does not propagate (there is no $\partial_0 A^0$ term in the lagrangian, unless it is introduced via the gauge condition). Therefore $A^0$ is determined by the positions of the charges, separately for each component of the wave function and at each instant of time. Imposing a non-vanishing boundary condition at spatial infinity when solving for $A^0$ gives rise to the linear term in the potential \eq{potential}, with $k$ determined by the asymptotic field strength. Thus the QCD scale $\lqcd$ may appear in the solution of the equations of motion even though the equations themselves have no such parameter.

The possibility of studying the properties of relativistic bound states within the very constrained framework of perturbation theory is interesting in its own right.  Bound states of lowest order in $\alpha$ and $\hbar$ retain the Lorentz and gauge invariance of the theory. The dynamical boost invariance of the equal-time bound states found using this method is non-trivial and (to my knowledge) unique. In considering the physical relevance of such an approach it is worth recalling that we rarely reflect on the reason (in fact we do not know) why QED does {\em not} confine electric charge. This ignorance has not kept us from doing perturbative calculations, implicitly assuming $A^0$ to vanish at large distances. The validity of this expansion for QED is certified by its precise agreement with data.

\section{Retarded vs. Feynman propagation \label{retprop}}

Tree (or Born) diagrams usually give a good first approximation to scattering amplitudes. They are special in the sense that all internal propagators are off-shell and hence insensitive to the $\ieps$ prescription of the propagators. Specifically, Feynman and retarded fermion propagators 
\beq\label{sfr}
S_{F/R}(p^0,\pv) = i\frac{\slashed{p}+m_{e}}{(p^0-E_{p}+\ieps)(p^0+E_{p}\mp\ieps)}
\eeq
give the same result for Born terms, whereas Feynman propagators are required in loop integrals.
The choice of propagator nevertheless makes a difference even at Born level when the interactions are time-ordered ($p^0 \to t$). The $E<0$ components propagate backward in time when using the Feynman prescription,
\beq\label{sft}
S_{F}(t,\pv) = \frac{1}{2E_{p}}\left[\theta(t)(E_{p}\gamma^0-\pv\cdot\gv+m_{e})e^{-iE_{p}t} -\theta(-t)(E_{p}\gamma^0+\pv\cdot\gv-m_{e})e^{iE_{p}t}\right]
\eeq
which gives rise to $Z$-diagrams as shown in \fig{fig2}. With the retarded prescription there is only forward propagation,
\beq\label{srt}
S_{R}(t,\pv) = \frac{\theta(t)}{2E_{p}}\left[(E_{p}\gamma^0-\pv\cdot\gv+m_{e})e^{-iE_{p}t} +(E_{p}\gamma^0+\pv\cdot\gv-m_{e})e^{iE_{p}t}\right]
\eeq
The first diagram on the {\em rhs.} of \fig{fig2} thus gives the entire time-ordered contribution when the retarded prescription is used, and the $Z$-contribution vanishes. For either prescription the sum of all time-ordered diagrams (integrated over the interaction times) gives the same Born amplitude.
%
\EPSFIGURE[hbt]{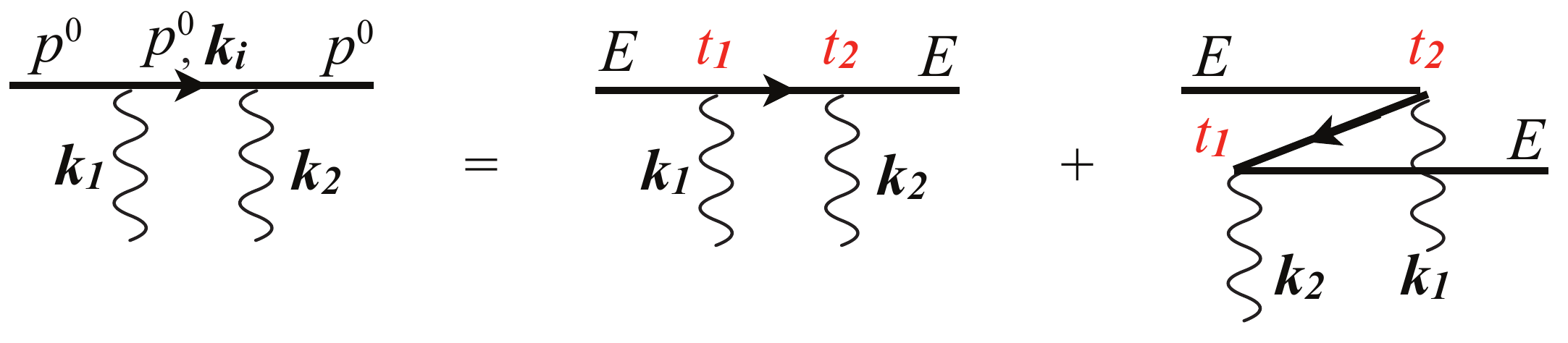,width=1.\columnwidth}{The tree diagram in 4-momentum $(p^0,\pv)$ space on the left splits into two time-ordered diagrams in $(t,\pv)$ space (right hand side) when using the Feynman propagator \eq{sft}. With the retarded propagator \eq{srt} both the positive and negative energy contributions are given by the first diagram on the right, and the $Z$-diagram is absent.\label{fig2}}
%

There is usually no need to time-order the interactions of scattering amplitudes, hence also the issue of $\ieps$ prescription is irrelevant at Born level. Time-ordering is, however, required to describe bound states using wave functions that give the spatial distribution of the constituents at an instant of time. When using Feynman propagators the infinite set of diagrams that generates a bound state pole has contributions with an unlimited number of nested $Z$-diagrams. This means that the wave function has components with any number of constituents.

%
\EPSFIGURE[hbt]{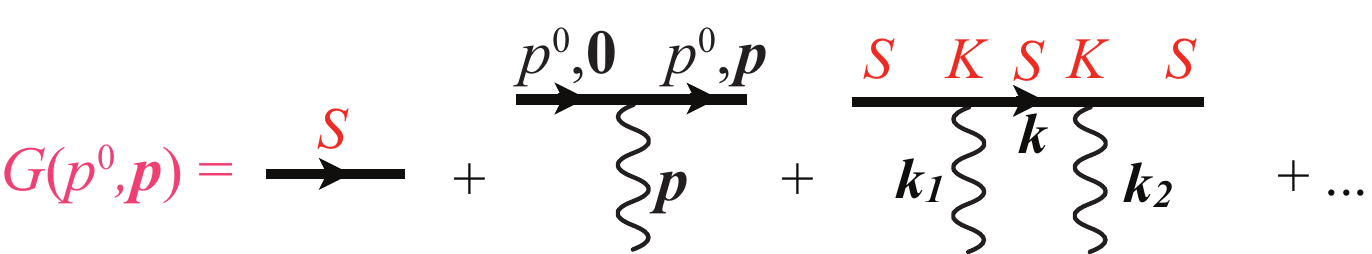,width=1.\columnwidth}{Electron scattering from a static external potential. The $p^0$ component of the electron momentum does not change during the scattering. The initial and final electron momenta are denoted $(p^0,\bs{0})$ and $(p^0,\pv)$, respectively.\label{fig3}}
%

As an illustration consider an electron in a static Coulomb field (\fig{fig3}). The energy component $p^0$ of the electron's 4-momentum is unchanged by the interactions, which only transfer 3-momentum. In the absence of loop corrections the electron's Green function is given by the sum 
\beq\label{dseq}
G(p^0,\pv) = S+SKS+SKSKS+\ldots = S+SKG = \frac{R(E_R,\pv)}{p^0-E_R}+\ldots
\eeq
where $S$ is the free propagator and $K$ the Coulomb interaction kernel $-ie\gamma^0 A^0(\kv)$ (the product involves a convolution over $\kv$).
The last equality in \eq{dseq} displays a bound state pole contribution. It is easy to verify that the residue $R(E_R,\pv)$ satisfies the Dirac equation in momentum space, \ie, we have Dirac bound states.

For $p^0>0$ the electron propagators $S$ in each diagram of \fig{fig3} are insensitive to the $\ieps$ prescription at the negative energy pole since $p^0 \neq -E_p$ in \eq{sfr}. Consequently the positions $p^0=E_R$ of the bound state poles in $G(p^0,\pv)$ are exactly the same whether we use $S_F$ or $S_R$ propagators. Yet if we time-order the interaction vertices these two propagators will generate very different diagrams. Each Feynman propagator gives a $Z$-contribution as in \fig{fig2}, while with the retarded propagator only the single electron (of positive or negative energy) is present at any intermediate time. The equal-time wave functions of the bound state are thus quite different in the two cases. The wave function we usually solve using the Dirac equation has only the single electron degree of freedom and corresponds to the use of a retarded electron propagator. The wave function of Dirac bound states corresponding to the Feynman propagator has  an unlimited number of Fock components and is seldom (ever?) given explicitly.

\section{The Dirac wave function \label{dirwf}}

The bound states $\ket{\varphi}$ of non-relativistic quantum mechanics are usually determined as eigenstates of the hamiltonian,
\beq\label{ham}
H\ket{\varphi}= E\ket{\varphi}
\eeq
The same condition holds in relativistic field theory, but is impractical even for the simple case of an electron bound by a Coulomb potential just discussed. Since the hamiltonian can create an $e^+e^-$ pair from the Coulomb field the state $\ket{\varphi}$ must necessarily contain an infinite number of pairs -- the same conclusion that we reached previously. Hence in relativistic theory one usually determines bound states as poles of Green functions in 4-momentum space rather than using a hamiltonian formulation.

According to our discussion above the standard (single particle) Dirac wave function follows from using \emph{retarded} propagators in a perturbative evaluation of Green functions as in \eq{dseq}. We may then ask whether we can set the boundary conditions in the operator equation \eq{ham} correspondingly, such that it defines a single particle bound state $\ket{\varphi}$ with a Dirac wave function. For this we need a ``retarded vacuum'' in which the electron propagator
\beq\label{sr2}
S_{R}(x-y)= {_{R}\bra{0}}\,T[\psi(x)\bar\psi(y)]\,\ket{0}_{R}
\eeq
agrees with the retarded one in \eqref{srt}. The definition
\beq\label{retvac}
\ket{0}_{R} = N^{-1}\prod_{\pv,\lambda} d_{\pv,\lambda}^\dag \ket{0}
\eeq
where all the positron states are filled\footnote{Equivalently, in the retarded vacuum all the negative energy states are empty, which is why pair production is suppressed.} and $N$ is an (infinite) normalization constant works since 
\beq\label{psiann}
\psi(x)\ket{0}_{R}=0
\eeq
implies no contribution for $x^0 < y^0$ in \eq{sr2}.
A single-electron state with both positive and negative energy components can then be parametrized by a Dirac (c-number spinor) wave function $\varphi(x)$ as
\beq\label{dstate}
\ket{\varphi,t} = \int d^3\xv\, \psi_{\alpha}^\dag(t,\xv)\varphi_{\alpha}(\xv)\ket{0}_{R}
\eeq
where a sum over the Dirac index $\alpha$ is implied. With the QED hamiltonian in the Interaction Picture
\beq\label{ipham}
H(t) = \int d^3\xv\, \bar\psi(t,\xv)\big[-i\nabv\cdot\gv+m+e\gamma^0 A^0(\xv)\big]\,\psi(t,\xv)
\eeq
the state \eq{dstate} in \eq{ham} gives, using $\big\{\psi_{\alpha}(t,\xv),\psi_{\beta}^\dag(t,\xv')\big\}=\delta^3(\xv-\xv')\,\delta_{\alpha\beta}$ and \eq{psiann}, the Dirac equation for the wave function $\varphi(\xv)$ of a bound state of energy $E$ in the external Coulomb potential $A^0(\xv)$,
\beq\label{diraceq}
(-i\nabv\cdot\gv+e\gamma^0 A^0(\xv) +m)\varphi(\xv) = E\gamma^0 \varphi(\xv)
\eeq

In Sec. \ref{retprop} we saw that the energy $E$ is independent of the boundary condition (Feynman or retarded) when the potential is static and there are no loop corrections. The possibility to describe the same state $\ket{\varphi}$ using different wave functions is not so surprising when we recall that the time-ordering of events which are separated by a space-like distance depends on the observer. Thus the relative magnitude of the two time-ordered diagrams in \fig{fig2}  depends on the frame, only their sum is Lorentz-invariant\footnote{Provided that also the external potential is Lorentz-transformed. Lorentz covariance is more properly a feature of states that are bound by the mutual interactions of their constituents and experience no external force. I consider this case in the next Section.}. In the infinite momentum frame the $Z$-diagram vanishes altogether and one arrives at a limiting picture where the hamiltonian does not create particles from the perturbative vacuum. Here I propose to eliminate the $Z$-diagrams in a different way, by using retarded fermion propagators at Born level. The bound states are found to be Lorentz-covariant and can thus be boosted to an arbitrary frame. To this lowest order contribution perturbative corrections may then be systematically applied.

\section{Dynamical boost invariance\label{dynboost}}

Many approaches to relativistic bound states strive to retain manifest Lorentz covariance. A two-particle Fock state wave function for a bound state $\ket{\varphi,P}$ with CM momentum $P$ is covariantly defined by
\beq
\Phi_{\alpha\beta}^{P}(x-y)\equiv \bra{0}{\rm T}\big\{\psi_{\alpha}(x)\bar\psi_{\beta}(y)\}\ket{\varphi,P}
\eeq
This Bethe-Salpeter wave function has a `kinematical' frame dependence,
\beq\label{lorrel}
\Phi^{P'}(x'-y')=S(\Lambda)\Phi^{P}(x-y)S^{-1}(\Lambda)
\eeq
where $S(\Lambda)$ is the usual Dirac matrix representation of the Lorentz transformation $\Lambda$, $P'=\Lambda P$ and $x'-y'=\Lambda(x-y)$. Eq. (\ref{lorrel}) relates a Bethe-Salpeter wave function defined at equal time ($x^0=y^0$) to wave functions which generally are at unequal time (${x'}^0\neq {y'}^0$). Equal-time wave functions are related to each other in a non-trivial way since the notion of equal time is frame-dependent.

Quantum field theory is generally not manifestly covariant since one quantizes on a space-like surface, usually the one of equal time \cite{Dirac:1949cp}. Poincar\'e invariance is ensured by requiring the transformation generators to satisfy the corresponding Lie algebra. Generators which preserve the quantization plane (such as space translations and rotations for equal-time surfaces) define `kinematic' transformations for which the symmetry is manifest. Generators which transform the quantization plane (such as time translation and boosts for equal-time surfaces) involve interactions and correspond to `dynamical', hidden symmetries.

A distinguishing feature of bound states is that they have a stationary time dependence, \ie, they are eigenstates of the hamiltonian as in \eq{ham}. Their wave functions define the spatial distribution of the constituents at an instant of time. Boosting equal-time states is a dynamical transformation which is generally as difficult as diagonalizing the hamiltonian. This and the problem of unlimited particle production discussed above challenge hamiltonian approaches to relativistic bound states.

Classical relativity would suggest that equal-time wave functions Lorentz contract like a rigid rod, which is indeed often claimed without further discussion. This issue was only addressed recently in the basic case of the Hydrogen atom \cite{Jarvinen:2004pi}. It turned out that the wave function of the moving atom receives a contribution from both the $e^-p$ and the $e^- p\gamma$ Fock states. In the atomic rest frame the instantaneous Coulomb field $A^0$ dominates and only the electron and proton constituents are present. When these charged constituents move relativistically with the CM they generate a propagating vector field $\bs{A}$ implying intermediate states containing (longitudinal) photons. The $e^- p\gamma$ component of the atomic wave function is required for the energy eigenvalue of the Hydrogen atom to have the correct dependence on its CM momentum, $E=\sqrt{\bs{P}^2+M^2}$, but does not Lorentz contract according to classical relativity.

In the previous section we saw how a relativistic electron bound by a static Coulomb potential is described by the single particle Dirac wave function when one uses retarded propagators. The bound state energies are insensitive to the choice of propagator when only tree diagrams are retained, which we referred to as the Born approximation. Provided this approximation indeed defines the lowest order in an $\hbar$ expansion it must share all the symmetries, and in particular dynamical boost invariance, with the exact result. Hence we turn now to a discussion of the $\hbar$ expansion in field theory.

\section{The $\hbar$ expansion\label{hbarsect}}

It is generally accepted that each loop in a Feynman diagram brings a factor of $\hbar$ \cite{Itzykson:1980rh,Iliopoulos:1974ur}, and that physics in the limit of $\hbar \to 0$ is classical. However, the first statement has been challenged \cite{Holstein:2004dn}, and the second statement also needs to be stated more precisely. Classical quantities such as mass and charge appear, for dimensional reasons, multiplied by powers of $\hbar$ in the lagrangian. In order to properly define an $\hbar \to 0$ limit one must therefore specify the behaviour of all parameters in the theory \cite{Brodsky:2010zk}. 

It is instructive to consider the $\hbar \to 0$ limit of the harmonic oscillator in quantum mechanics.
The propagation of a particle from $(t_{i},x_{i})$ to $(t_{f},x_{f})$ is given by the path integral
\beqa\label{ho}
\mA(x_{i},x_{f};t_{f}-t_{i}) &=& \int[\mD x(t)]\exp\left[\frac{i m}{2\hbar}\int_{t_{i}}^{t_{f}}dt ({\dot x}^2-\omega^2 x^2)\right] \nn\\[2mm]
 &=& \int[\mD\xi(t)]\exp\left[\frac{im}{2}\int_{t_{i}}^{t_{f}}dt ({\dot \xi}^2-\omega^2 \xi^2)\right]
\eeqa
In the second expression all explicit dependence on $\hbar$ was removed by a mere redefinition of the variables, $\xi\equiv x/\sqrt{\hbar}$. Hence it is clear that the full quantum structure of the harmonic oscillator, including its bound states, remains in the $\hbar \to 0$ limit. The usual argument for an approach to classical physics is that the rapid variation of the phase factor $\exp(iS/\hbar)$ selects the classical path for which the action $S$ is stationary. This constraint is evaded for paths whose length scales as $x \propto \sqrt{\hbar}$ since then $S \propto \hbar$. The bound state wave functions are in this category since their extent scales with $\xi$. Classical physics dominates in the $\hbar \to 0$ limit only for paths whose length is independent of $\hbar$.

In field theory one usually sets $\hbar=1$ at the outset, causing the dimensions of length $L$ and energy $E$ to be inversely related. For the present discussion \cite{Brodsky:2010zk} we wish to keep the factors of $\hbar$ explicit, with dimension $[\hbar] = E\cdot L$. We keep $c=1$, so the units of space and time are both $L$. The QED action
\beq\label{qedact}
{\cal S}_{QED} =
\int d^4x\left[\bar\psi\left(i\slashed{\partial}-\frac{e}{\hbar}\slashed{A}-\frac{m}{\hbar}\right)\psi-\inv{4}F_{\mu\nu}F^{\mu\nu}\right]
\eeq
should have the same dimension as $\hbar$, hence $[\psi]=E^{1/2}L^{-1}$ and $[A]=E^{1/2}L^{-1/2}$. From the fact that the fine structure constant $\alpha=e^2/(4\pi\hbar)$ is dimensionless we conclude that the dimension of the classical charge is $[e]=E^{1/2}L^{1/2}$. This is why the charge in the QED action \eq{qedact} enters in the form $e/\hbar$. Similarly the classical mass appears as $m/\hbar$.

In considering the classical limit of quantum field theory it seems natural to keep the classical quantities $e$ and $m$ fixed as $\hbar  \to 0$ \cite{Holstein:2004dn}. For a connection between the power of $\hbar$ and the number of loops we need instead to keep $\tilde e \equiv e/\hbar$ and $\tilde m \equiv m/\hbar$ fixed \cite{Itzykson:1980rh}. After a rescaling of the fields, $\tilde{\psi}\equiv \psi/\sqrt{\hbar}$ and $\tilde{A}\equiv A/\sqrt{\hbar}$ we get

\beq\label{qedres}
\inv{\hbar}{\cal S}_{QED} = 
\int d^4x\left[\tilde{\bar\psi}\left(i\slashed{\partial}-\tilde{e}\sqrt{\hbar}\tilde{\slashed{A}}-\tilde{m}\right)\tilde{\psi}\right]
\eeq
where $\hbar$ appears exclusively in the combination $\tilde{e}\sqrt{\hbar}$. Adding a loop to any given Feynman diagram gives a factor $(\tilde{e}\sqrt{\hbar})^2$ and thus a factor of $\hbar$. This establishes the equivalence between the $\hbar$ and loop expansions\footnote{A Green function also has a factor $\hbar^{1/2}$ for each of its external legs, due to the conversions $\psi\to\tilde{\psi}$ and $A\to\tilde{A}$ of the corresponding operators. Thus the free electron and photon propagators are of \order{\hbar}.}. The conclusion is the same for QCD. 

%
\EPSFIGURE[hbt]{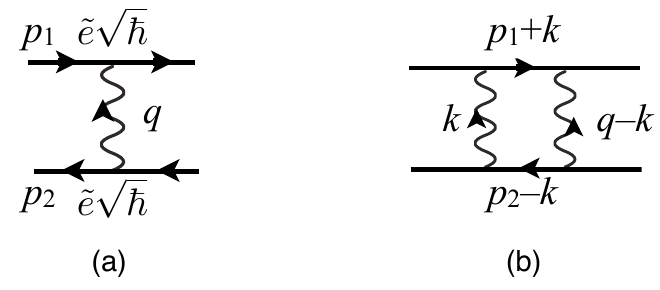,width=.6\columnwidth}{The first two ladder diagrams contributing to non-relativistic atoms in the limit of small coupling $\alpha$.\label{fig4}}
%

How about the $\hbar$ expansion for bound states? In the weak coupling limit ($\alpha \to 0$) QED atoms are given by a sum of ladder diagrams, the first two of which are shown in \fig{fig4}. Each ladder increases the number of loops by one, so atoms would seem to be of no definite order in $\hbar$. On the other hand, the binding energies $E_n = - \halft \alpha^2 m_e/n^2$ are of definite power of $\alpha={\tilde e}^2\hbar/4\pi$. The loop and $\hbar$ expansions appear not to be equivalent in this case.

Bound states are indeed a special case because the momenta in the ladders depend on $\alpha$.  For the ladder sum to maintain an overlap with the bound state wave function the momentum transfer in \fig{fig4} must be of the order of the Bohr momentum, $|\qv| = \morder{\alpha m_e}$, while the energy transfer should be commensurate with the binding energy, $q^0 = \morder{\alpha^2 m_e}$. Hence the propagators can contribute negative powers of $\alpha$. For diagram (a) in \fig{fig4} we have $\mA(a) \sim \alpha/\qv^2 \propto 1/\alpha$. In diagram (b) the four vertices and the $dk^0d^3\kv$ of the loop integral measure give $\alpha^2\cdot \alpha^2\cdot \alpha^3=\alpha^7$ in the numerator. The fermion propagators are each off-shell by the binding energy and hence $\propto \alpha^{-2}$, while each photon propagator $\propto \alpha^{-2}$ as in $\mA(a)$. Thus the four propagators in (b) contribute $\propto \alpha^{-8}$, implying that $\mA(b)\propto \alpha^{-1}$ is of the same order in $\alpha$ and $\hbar$ as $\mA(a)$. This also explains why the ladder sum can diverge at the bound state energies, however small is $\alpha$. The ladder diagrams are enhanced ($\propto 1/\alpha$) in the weak coupling limit, and thus give the leading contribution to the bound state poles. In diagrams with loop corrections to vertices or propagators the loop momentum does not decrease with $\alpha$, hence such diagrams give true higher order corrections in $\alpha$ and $\hbar$.

As $\alpha \to 0$ the sum of ladder diagrams in \fig{fig4} describes non-relativistic scattering from a static Coulomb potential as shown in \fig{fig3}. Hence the Schr\"odinger equation gives the Born contribution, of lowest order in $\hbar$, to atomic bound states. The Dirac equation with a Coulomb potential $A^0=-\alpha/r$ can also be obtained from the sum of ladder diagrams (with crossed ladders included) in the limit where one of the fermion masses in \fig{fig4} grows large \cite{Brodsky:1971sk}. The relativistic features which distinguish the Dirac from the Schr\"odinger bound states are of the same  \order{\alpha} as loop contributions. Thus the Dirac equation for the Hydrogen atom contains only a subset of the relativistic corrections to the Schr\"odinger equation \cite{Itzykson:1980rh}.

\section{$A^0$ from Gauss' law \label{gausssect}}

Having seen that the Schr\"odinger equation defines a Born term for non-relativistic atoms I now return to the hamiltonian field theory formulation \eq{ham} of bound states. Previously I assumed a fixed external potential, now we shall recall how the Coulomb potential arises from the equations of motion. I take the constituents to have distinct flavor, an electron $e^-$ and a muon $\mu^+$ for definiteness\footnote{The annihilation contributions of an $e^+e^-$ bound state are in any case of higher order in $\alpha$ than I shall be concerned with here.}.

In Coulomb gauge $\nabv\cdot\bs{A}=0$ the QED lagrangian
\beq\label{lqed}
{\cal L} = -\inv{4}F_{\mu\nu}F^{\mu\nu} + \sum_{f=e,\mu}\bar\psi_f(i\slashed{\partial}-e\slashed{A}-m)\psi_f
\eeq
defines the operator constraint 
\beq\label{gauss}
-\nabv^2 A^0(t,\xv) = e\sum_{f=e,\mu}\psi_f^\dag(t,\xv)\psi_f(t,\xv)
\eeq
Since there is no time derivative $\partial_0 A^0$  we may express $A^0$ in terms of the fermion operators at each instant of time,
\beq\label{asol}
A^0(t,\xv)=\int d^3\yv\,\frac{e}{4\pi|\xv-\yv|}\sum_{f=e,\mu}\psi_f^\dag(t,\yv)\psi_f(t,\yv)
\eeq
This allows $A^0$ to be eliminated from the action. From $-F_{\mu\nu}F^{\mu\nu}/4$ we get
\beq\label{actf}
\inv{2}\int d^3\xv (\nabv A^0)^2 = \inv{2}\int d^3\xv A^0(-\nabv^2A^0)
= \inv{2}\int d^3\xv\,\sum_f\psi_f^\dag\,eA^0\,\psi_f
\eeq
which cancels half the Coulomb interaction of the fermions. The lagrangian becomes
\beq\label{lqed0}
{\cal L} = \inv{2}(\partial_0 \bs{A})^2 -\inv{4}F_{ij}F^{ij} + \sum_{f=e,\mu}\bar\psi_f\big(i\slashed{\partial}-\halft e\gamma^0 A^0 +e\gv\cdot\bs{A}-m_f\big)\psi_f
\eeq
with $A^0$ given by \eq{asol}. So far we made no approximations.

The $e^-\mu^+$ bound state may be expressed in analogy to the single particle state \eq{dstate} as
\beq\label{ffqed}
\ket{E,t}=\int d^3\yv_{1}d^3\yv_{2}\,\psi_{e}^{\dag}(t,\yv_{1})\chi(\yv_{1}-\yv_{2})\psi_{\mu}(t,\yv_{2})\ket{0}
\eeq
where the rest frame wave function $\chi$ depends only on the difference of the fermion coordinates.  For small $\alpha=e^2/4\pi$ the state is non-relativistic and the $4\times 4$ matrix $\chi$ is dominated by its large (``upper'') components. Consequently the couplings of the vector components $\bs{A}$ of the photon field to the fermions are suppressed, and we may at lowest order in $\alpha$ neglect them in the hamiltonian,
\beq\label{ipham2}
H(t) = \int d^3\xv\sum_{f=e,\mu} \bar\psi_f(t,\xv)\big[-i\nabv\cdot\gv+m+\halft e\gamma^0 A^0(\xv)\big]\,\psi_f(t,\xv)
\eeq
The non-local four-fermion interaction term which arises with the expression \eq{asol} for $A^0$ looks forbidding, but is harmless in the non-relativistic case where particle production is neglected. The only contribution to the eigenvalue equation \eq{ham} then arises from $\psi_e$ in the hamiltonian annihilating $\psi^\dag_e$ in the state \eq{ffqed} through $\big\{\psi_{\alpha}(t,\xv),\psi_{\beta}^\dag(t,\yv)\big\}=\delta^3(\xv-\yv)\,\delta_{\alpha\beta}$, and similarly for the muon fields. Keeping only the two-particle contribution the eigenvalue condition becomes
\beqa\label{ham2}
H(t)\ket{E,t} &=& \int d^3\yv_{1}d^3\yv_{2}\,\psi_{e}^{\dag}(t,\yv_{1})\bigg[\gamma^0(-i\overrightarrow{\nabv}_1\cdot\gv+m_e)\chi(\yv_{1}-\yv_{2})\nn\\
&-&\chi(\yv_{1}-\yv_{2})\gamma^0(i\overleftarrow{\nabv}_2+m_\mu)-\frac{\alpha}{|\yv_1-\yv_2|}\bigg]\psi_{\mu}(t,\yv_{2})\ket{0}
\eeqa
Requiring the coefficient of $\psi_{e}^{\dag}(t,\yv_{1})\psi_{\mu}(t,\yv_{2})$ to agree with that of $E\ket{E,t}$ implies
\beq\label{bses}
\gamma^0(-i\overrightarrow{\nabv}\cdot\gv+m_e)\chi(\yv)-\chi(\yv)\gamma^0(-i\overleftarrow{\nabv}\cdot\gv+m_\mu)
= [E-V(|\yv|)]\chi(\yv)
\eeq
where 
\beq\label{coulpot}
V(|\yv|) = -\frac{\alpha}{|\yv|}
\eeq
is the standard Coulomb potential. In the weak coupling limit the bound state equation \eq{bses} reduces \cite{Hoyer:2009ep} to the Schr\"odinger equation
\beq
\left(-\frac{\nabv^2}{2m_e}-\frac{\nabv^2}{2m_\mu}+V\right)\chi=(E-m_e-m_\mu)\chi
\eeq
In this section I recalled that the $A^0$ potential can be expressed in terms of the charged fields as in \eq{asol}. I next discuss how this expression may be modified by a homogeneous solution of the field equations, corresponding to a different boundary condition. This gives a linear potential and thus relativistic bound states even at lowest order in $\alpha$.

\section{A confining boundary condition on $A^0$ \label{confbound}}

The strong interactions are characterized by a scale $\lqcd \sim 200$ MeV which is not present in the classical QCD lagrangian. The scale is introduced when loop corrections are renormalized, resulting in a ``dimensional transmutation'' where the fixed dimensionless coupling is replaced by the running $\as(Q)$. $\lqcd$ also determines the confinement scale through vacuum expectation values such as $\bra{0}\as F_{\mu\nu}F^{\mu\nu}\ket{0} \propto \lqcd^4$. It is worth recalling that even the QED vacuum is very complicated since the hamiltonian creates and destroys particles, implying an infinite number of balancing fluctuations. Nevertheless, in QED one successfully expands around the empty, perturbative vacuum and includes vacuum fluctuations only perturbatively. 

The QCD scale is introduced phenomenologically in the quark model, by postulating a linear potential with scale $k$ in \eq{potential}. Can there be an approach within perturbative QCD that allows to arrive at a quark model picture of hadrons? As I discussed in Sec. \ref{hbarsect} the Schr\"odinger equation represents a Born level approximation, \ie, it does not include loop effects. How could the scale $\lqcd$ appear at Born level in QCD? One possibility (and apparently the only one) is via a boundary condition on the color field strength. Gauss' law \eq{gauss} (or rather its QCD equivalent) is particularly relevant since it concerns the instantaneous field $A^0$. Interactions transmitted by the propagating gluon fields $\Av$ necessarily imply that hadrons have gluons in their Fock states, rather than being (at Born level) $\qu\qb$ and $\qu\qu\qu$ states as in the quark model. The fact that the spectrum of heavy quarkonia is qualitatively similar to the atomic spectrum furthermore suggests that they bind similarly, through the Coulomb potential $A^0$.

Consider adding a homogeneous solution of Gauss' constraint \eq{gauss} to the standard expression \eq{asol} (still using QED as illustration),
\beq\label{asol2}
A^0(t,\xv)=\int d^3\yv\,\frac{e}{4\pi|\xv-\yv|}\sum_{f=e,\mu}\psi_f^\dag(t,\yv)\psi_f(t,\yv)+\la^2\chat\cdot\xv
\eeq
where $\la$ and the unit vector $\chat$ are independent of $\xv$. This contribution implies a constant energy density $\halft (\nabv A^0)^2 = \halft\la^2$ at asymptotic $\xv$. The possibility to add this term means that it preserves the stationarity of the action under {\em local} variations of $A^0$, but not necessarily stationarity under the {\em global} variation of $\chat$. Keeping only the two-particle $\ket{\yv_{1},\yv_{2}}\equiv \psi_{e}^{\dag}(t,\yv_{1})\psi_{\mu}(t,\yv_{2})\ket{0}$ Fock state contribution,
\beq
A^0(t,\xv)\ket{\yv_{1},\yv_{2}}
=\left[\frac{e}{4\pi}\left(\inv{|\xv-\yv_{1}|}-\inv{|\xv-\yv_{2}|}\right)+\la^2\chat\cdot\xv\right]\ket{\yv_{1},\yv_{2}}
\eeq
The electron and muon  generate an electric dipole field, which results in the standard $1/r$ potential in \eq{bses} (when the infinite, $\yv_{1},\yv_{2}\,$-independent contributions are discarded). The instantaneous $A^0$-field is thus {\em specific for each position of the fermions}. The stationarity of the action under the global variation of $\chat$ should similarly be ensured separately for each Fock state. The field strength contribution may be evaluated as in \eqref{actf} through partial integration (except for  the $\la^4$ term). This gives to \order{e},
\beq\label{actlin}
\inv{2}\int d^3\xv\,\Big[\la^4-e\sum_f\psi_f^\dag(x) \la^2\chat\cdot\xv\, \psi_f(x)\Big]\ket{\yv_{1},\yv_{2}} = \inv{2}\Big[\la^4\int d^3\xv\,- e\la^2\,\chat\cdot(\yv_{1}-\yv_{2})\Big]\ket{\yv_{1},\yv_{2}}
\eeq
The first term on the \rhs is proportional to the volume of space and is due to the constant energy density $\halft(\nabv A^0)^2$. In bound state calculations this infinite term can be discarded provided $\la$ is independent of $\yv_{1}$ and $\yv_{2}$. Stationarity of the second term \wrt variations of $\chat$ imposes (up to a sign) 
\beq\label{chatqed}
\chat =  \frac{\yv_{1}-\yv_{2}}{|\yv_{1}-\yv_{2}|}
\eeq
which gives a linear potential contribution to the interaction energy in \eq{ipham2},
\beq\label{linpol}
V(\yv_{1},\yv_{2}) =  \halft e\la^2\,|\yv_{1}-\yv_{2}|
\eeq

Assuming a non-vanishing energy density ($\la\neq 0$) thus leads to a linear potential. It is interesting to note that this is consistent with translation invariance only for \emph{neutral} states. For an electron with charge $e_{1}$ and a muon with charge $e_{2}$ we would have obtained a potential $V\propto |e_{1}\yv_{1}-e_{2}\yv_{2}|$, which is not invariant under a translation $\yv_{i} \to \yv_{i}+\bs{c}$. The restriction to neutral bound states is consistent with color confinement in QCD.

The linear potential \eqref{linpol} is of \order{e} and thus leading \wrt the \order{e^2} photon exchange contribution \eqref{coulpot}. This allows to study bound states self-consistently using only the linear potential, and introducing photon exchange perturbatively. For relativistic states also the propagating components $\Av$ of the photon field would contribute at \order{e^2}. As I mentioned in Sec. 6 the longitudinal components of the photon field must be taken into account even for the Hydrogen atom when it is in relativistic CM motion. 

In the previous discussion I considered only two-particle Fock states, as appropriate for non-relativistic $e^-\mu^+$ bound states. This restriction can be  removed using the observation in Sec.~5 that pair production for Dirac states is ``hidden'' when using retarded boundary conditions at asymptotic times. At the Born level, \ie, at lowest order in $\hbar$, there are no loops and the energy eigenvalues are unaffected by the $\ieps$ prescription of the propagators. It should be recalled that despite the apparent simplicity of the ensuing, Dirac-type ``valence'' wave functions the relativistic bound states actually contain an infinite number of pairs when Feynman boundary conditions are imposed at $t=\pm\infty$. 

The generalization of the ``retarded vacuum'' \eqref{retvac} to the case of two fermions $e^-$ and $\mu^+$ is straightforward,
\beq\label{retvac2}
\ket{0}_{R} = N^{-1}\prod_{\pv,\lambda} d_{e}^{\dag}(\pv,\lambda)\, b_{\mu}^{\dag}(\pv,\lambda) \ket{0}
\eeq
implying
\beq\label{psiann2}
\psi_{e}(x)\ket{0}_{R}=\psi_{\mu}^\dag(x)\ket{0}_{R}=0
\eeq
When the $\ket{E,t}$ bound state in \eqref{ffqed} is defined using the retarded $\ket{0}_{R}$ rather than the perturbative vacuum $\ket{0}$ the property \eqref{psiann2} ensures that only two-particle states contribute.

\section{$\qu\qb$ states in QCD\label{mesons}}

I used QED in the above discussion to illustrate the essential features of the approach. The method of deriving bound states at Born level in $\hbar$ and at lowest order in the coupling can also be applied to QCD, with
\beqa\label{qcdlag}
{\cal L}_{QCD} &=& -\quart F_{a}^{\mu\nu}F_{\mu\nu}^{a} + \sum_{f=u,d}\bar\psi_{f}^{A}(i\slashed{\partial}-g\slashed{A}_{a}T_{a}^{AB}-m_{f})\psi_{f}^{B} \nn\\
F_{a}^{\mu\nu} &=& \partial^\mu A_{a}^\nu - \partial^\nu A_{a}^\mu - gf_{abc}A_{b}^\mu A_{c}^\nu
\eeqa
The retarded vacuum now includes a product over color to make it gauge invariant,
\beq\label{retvac3}
\ket{0}_{R} = N^{-1}\prod_{\pv,\lambda,A} d_{u}^{A\dag}(\pv,\lambda)\, b_{d}^{A\dag}(\pv,\lambda) \ket{0}
\eeq
and the $u\bar d$ meson state is
\beq\label{ffqcd}
\ket{E,t}=\sum_{A,B}\int d^3\yv_{1}d^3\yv_{2}\,\psi_{u}^{A\dag}(t,\yv_{1})\chi^{AB}(\yv_{1},\yv_{2})\psi_{d}^{B}(t,\yv_{2})\ket{0}_{R}
\eeq
This state is invariant under (time-independent) gauge transformations $\psi(t,\xv) \to U(\xv)\psi(t,\xv)$ provided that the wave function is transformed as
\beq
\chi(\yv_{1},\yv_{2}) \to U(\yv_{1})\chi(\yv_{1},\yv_{2}) U^{\dag}(\yv_{2})
\eeq
We may then look for stationary states of the form \eqref{ffqcd} in a Coulomb gauge where the wave function has the standard ``color singlet'' form,
\beq\label{chising}
\chi^{AB}(\yv_{1},\yv_{2}) = \delta^{AB} \chi(\yv_{1},\yv_{2})
\eeq

The equation of motion for the gluon field
\beq\label{gluoneom}
\frac{\delta}{\delta A_a^\rho}{\cal L}_{QCD} = \big(\partial^\mu\delta_{ab}+gf_{abc}A_c^\mu\big)F_{\mu\rho}^b- g\sum_f\bar\psi_f^A\gamma_\rho T_a^{AB}\psi_f^B = 0
\eeq
has the perturbative solution
\beqa\label{asolqcd}
A_a^0(t,\xv)&=&\la_a^2\chat_a\cdot\xv+\int d^3\yv\,\frac{g}{4\pi|\xv-\yv|}\sum_{f}{\psi_f^A}^\dag(t,\yv)T_a^{AB}\psi_f^B(t,\yv)+\morder{g^2}\nn\\
\Av_a(t,\xv)&=&\morder{g}
\eeqa
provided $\la_a=0$ for $a\neq 3,8$. Introducing a homogeneous solution only for the color diagonal generators $T_3$ and $T_8$ is necessary to ensure the color structure \eq{chising} of the wave function and implies that the non-abelian terms in \eq{gluoneom} are of \order{g^2}.

For the solution \eq{asolqcd} the interaction part of the Lagrange function $L_{I}(t)=\int d^3\xv\, {\cal L}_{I,QCD}$ is
\beq\label{lagfun}
L_{I}(t)=\inv{2}\int d^3\xv \sum_a \Big[\la_a^4-g\la_a^2\chat_a\cdot\xv\sum_f{\psi_f^A}^\dag(t,\xv)T_a^{AB}\psi_f^B(t,\xv)+\morder{g^2}\Big]
\eeq
The (infinite) energy proportional to the volume of space must be the same for all Fock components of the bound state \eq{ffqcd}, which requires that the field strength
\beq\label{lamcond}
\la^4 \equiv \sum_{a=3,8}\la_a^4
\eeq
is a universal constant. I do not show this term in the following. The remaining part of $L_{I}(t)$ in \eq{lagfun} should be stationary on each Fock component $\ket{\yv_1,\yv_2;C}\equiv {\psi_u^C}^\dag(\yv_1){\psi_d^C}(\yv_2)\ket{0}_R$,
\beq\label{lagqcd}
L_{I}(t)\ket{\yv_1,\yv_2;C}=-\frac{g}{2}\sum_{a}\la_a^2\,T_a^{CC}\,\chat_a\cdot(\yv_1-\yv_2) \ket{\yv_1,\yv_2;C}+\morder{g^2}
\eeq
The variation \wrt the direction of $\chat_a$ imposes (up to a sign),
\beq\label{chatqcd}
\chat_a =  \frac{T_a^{CC}(\yv_{1}-\yv_{2})}{|T_a^{CC}(\yv_{1}-\yv_{2})|}
\eeq
The Lagrange function \eq{lagqcd} must be stationary also \wrt variations of the ratio $\la_3^2/\la_8^2$, under the constraint that $\la$ in \eq{lamcond} is fixed. Together with \eq{chatqcd} this gives
\beq\label{lagqcd2}
L_{I}(t)\ket{\yv_1,\yv_2;C}=-\frac{g\la^2}{2}\sqrt{\textstyle{\sum_{a}}\big(T_a^{CC}\big)^2}\,\big|\yv_1-\yv_2\big| \ket{\yv_1,\yv_2;C}+\morder{g^2}
\eeq
The SU(3) generator identity 
\beq\label{genid}
\textstyle{\sum_{a}}T_a^{AB}T_a^{CD}=\halft\big(\delta_{AD}\delta_{BC}-\thirt\delta_{AB}\delta_{CD}\big)
\eeq 
gives $\sum_{a}\big(T_a^{CC}\big)^2=\thirt$, ensuring that the eigenvalue in \eq{lagqcd2} is independent of the color $C$ of the quarks. For the interaction hamiltonian $H_I(t)=-L_{I}(t)+\morder{g^2}$ we have
\beq
H_I(t)\ket{\yv_1,\yv_2;C}=\frac{g\la^2}{2\sqrt{3}}\big|\yv_1-\yv_2\big|\, \ket{\yv_1,\yv_2;C}+\morder{g^2}
\eeq
The fact that the linear potential has the same strength for all color components of the wave function is consistent with the diagonal ansatz \eq{chising}.

The eigenvalue condition for the bound states defined by \eq{ffqcd} and \eq{chising} is then, up to terms of \order{g^2},
\beqa\label{qcdstat}
H(t)\ket{E,t}&=& \int d^3\xv d^3\yv_1 d^3\yv_2\sum_{f,A,B} \bar\psi_f^A(t,\xv)\Big[-i\nabv_\xv\cdot\gv+m_f+\frac{g\la^2}{2\sqrt{3}}\big|(\yv_1-\yv_2)\big|\gamma^0\,\Big]\,\psi_f^A(t,\xv)\nn\\[2mm]
&\times&\,\psi_{u}^{B\dag}(t,\yv_{1})\chi(\yv_{1},\yv_{2})\psi_{d}^{B}(t,\yv_{2})\ket{0}_{R} =E\,\ket{E,t}
\eeqa
Using
\beq
\big\{\psi_{f,\alpha}^A(t,\xv),\psi_{f',\beta}^{B\dag}(t,\yv)\big\}=\delta^3(\xv-\yv)\,\delta_{\alpha\beta}\,\delta_{ff'}\,\delta_{AB} \hspace{.5cm} {\rm and} \hspace{.5cm} \psi_u^A(x)\ket{0}_{R}={\psi_d^A}^\dag(x)\ket{0}_{R}=0
\eeq
we get the condition for the wave function $\chi(\yv_{1},\yv_{2})$ of \eq{chising},
\beq\label{bsemes}
\gamma^0(-i\overrightarrow{\nabv}_1\cdot\gv+m_u)\chi(\yv_{1},\yv_{2})-\chi(\yv_{1},\yv_{2})\gamma^0(i\overleftarrow{\nabv}_2+m_d)
= [E-V(|\yv_{1}-\yv_{2}|)]\,\chi(\yv_{1},\yv_{2})
\eeq
with the linear potential\footnote{The coefficient differs from that of Eq. (5.25) in \cite{Hoyer:2009ep} due to a different regularization of the infinite contribution $\propto$ the volume of space in \eq{lagfun}.}
\beq\label{linpot}
V(|\yv|) = \frac{g\la^2}{2\sqrt{3}}\,\big|\yv\big|+\morder{g^2}
\eeq

This bound state condition for the wave function is valid in any frame. For a state \eq{ffqcd} with CM momentum $\kv$ the wave function has the form 
\beq\label{cmsep}
\chi(\yv_{1},\yv_{2}) = e^{i\kv\cdot(\yv_{1}+\yv_{2})/2}\,\chi_{\kv}(\yv_{1}-\yv_{2})
\eeq
Substituting this in \eq{bsemes} gives
\beq\label{bsecm}
-i\nabv\cdot\com{\bs{\alpha}}{\chi_{\kv}(\yv)}+\halft\kv\cdot\acom{\bs{\alpha}}{\chi_{\kv}(\yv)}+m_{u}\gamma^0\chi_{\kv}(\yv)-\chi_{\kv}(\yv)\gamma^0 m_{d} = \big[E_k-V(|\yv|)\big]\chi_{\kv}(\yv)
\eeq
where $\bs{\alpha}=\gamma^0\gv$. As I discussed in Sec. \ref{dynboost}, equal-time wave functions transform dynamically under boosts. Thus \eq{bsecm} is not manifestly Lorentz covariant. Nevertheless, since we arrived at this equation through an expansion in the fundamental parameters $\hbar$ and $\as$, the energy eigenvalues should have the correct dependence on the CM momentum: $E_k=\sqrt{\kv^2+M^2}$. Remarkably, this relation is found to hold \cite{Hoyer:1985tz}. The wave function transforms dynamically under boosts, Lorentz contracting at a rate $\propto 1/(E-V)$ which depends on the potential, and hence on the separation $|\yv|$ between the quarks. 

The bound state equation \eq{bsecm} is (to my knowledge) the only case where a correct $\kv$-dependence of the energy has been obtained. This occurs only in the case of an exactly linear potential, as expected since the \order{g^2} gluon exchange corrections were neglected in the derivation.

\section{$\qu\qu\qu$ states in QCD\label{baryons}}

The derivation of $uds$ bound states (baryons) is similar to the above $\qu\qb$ case (mesons), but contains some new elements. The state is defined as
\beq\label{udsket}
\ket{E,t} = \sum_{A,B,C}\int d^3\yv_{1} d^3\yv_{2} d^3\yv_{3}\,\psi_{u\alpha_{1}}^{A\dag}(t,\yv_{1})\psi_{d\alpha_{2}}^{B\dag}(t,\yv_{2})\psi_{s\alpha_{3}}^{C\dag}(t,\yv_{3})\chi_{ABC}^{\alpha_{1}\alpha_{2}\alpha_{3}}(\yv_{1},\yv_{2},\yv_{3})\ket{0}_{R}
\eeq
where now
\beq\label{retvac4}
\ket{0}_{R} = N^{-1}\prod_{\pv,\lambda,A} d_{u}^{A\dag}(\pv,\lambda)\, d_{d}^{A\dag}(\pv,\lambda)\, d_{s}^{A\dag}(\pv,\lambda) \ket{0}
\eeq
The baryon state \eq{udsket} is invariant under time independent gauge transformations $\psi^{A}(t,\xv) \to U^{AA'}(\xv)\psi^{A'}(t,\xv)$ provided the wave function is transformed as
\beq
\chi_{ABC}(\xv_{1},\xv_{2},\xv_{3}) \to U^{AA'}(\xv_{1})U^{BB'}(\xv_{2})U^{CC'}(\xv_{3})\chi_{A'B'C'}(\xv_{1},\xv_{2},\xv_{3})
\eeq
I assume that there is a gauge where
\beq\label{colbar}
\chi_{ABC}^{\alpha_{1}\alpha_{2}\alpha_{3}}(\yv_{1},\yv_{2},\yv_{3}) = \epsilon_{ABC}\chi^{\alpha_{1}\alpha_{2}\alpha_{3}}(\yv_{1},\yv_{2},\yv_{3})
\eeq

The solution \eq{asolqcd} of the QCD equations of motion gives the Lagrange function \eq{lagfun} with the same divergent contribution as before which requires that $\la$ in \eq{lamcond} is universal (independent of the state). Applying the Lagrange function (minus the divergent term) to a specific Fock component of the bound state \eq{udsket} where, according to \eq{colbar} $A,B,C$ is some permutation of $1,2,3$, gives
\beqa\label{lagbar}
L(t)\ket{\yv_{1},\yv_{2},\yv_{3};ABC}&\equiv& L(t)\psi_{u}^{A\dag}(\yv_{1})\psi_{d}^{B\dag}(\yv_{2})\psi_{s}^{C\dag}(\yv_{3})\ket{0}_{R}\\[2mm]
&=&-\frac{g}{2}\sum_a \la_a^2\, \chat_a\cdot\big(T_a^{AA}\yv_1+T_a^{BB}\yv_2+T_a^{CC}\yv_3\big) \ket{\yv_{1},\yv_{2},\yv_{3};ABC}\nn
\eeqa
The condition of stationarity aligns $\chat_a$ with the vector it multiplies in \eq{lagbar}, hence
\beq\label{lagbar2}
L(t)\ket{\yv_{1},\yv_{2},\yv_{3};ABC}=-\frac{g}{2}\sum_a \la_a^2\, \big|T_a^{AA}\yv_1+T_a^{BB}\yv_2+T_a^{CC}\yv_3\big| \ket{\yv_{1},\yv_{2},\yv_{3};ABC}
\eeq
Under translations $\yv_i \to \yv_i+\bs{c}$ the vector $\Yv_a = T_a^{AA}\yv_1+T_a^{BB}\yv_2+T_a^{CC}\yv_3$ shifts by $(\tr{T_a})\,\bs{c}=0$. Hence the color structure \eq{colbar} ensures translation invariance.

The expression \eq{lagbar2} is extremal for $\la_3^2/\la_8^2=|\Yv_3|/|\Yv_8|$, which depends on the quark positions $\yv_i$. The interaction hamiltonian, \ie, the potential, is then
\beq
V(\yv_{1},\yv_{2},\yv_{3})=\frac{g\la^2}{2}\sqrt{\Yv_3^2+\Yv_8^2}
\eeq 
In order to be consistent with the color structure \eq{colbar} this potential should be independent of the specific color permutation $A,B,C$ of the quarks at position $\yv_1,\yv_2,\yv_3$. This may be seen as follows:
\beqa
\sum_a\Yv_a^2&=&\sum_a \Big[T_a^{AA}\yv_1+T_a^{BB}\yv_2-\big(T_a^{AA}+T_a^{BB}\big)\yv_3\Big]^2\\
&=& \sum_a \Big[\big(T_a^{AA}\big)^2(\yv_1-\yv_3)^2+\big(T_a^{BB}\big)^2(\yv_2-\yv_3)^2+2T_a^{AA}T_a^{BB}(\yv_1-\yv_3)\cdot(\yv_2-\yv_3)\Big]\nn
\eeqa
where according to \eq{genid} $\sum_a \big(T_a^{AA}\big)^2=\thirt$ and $\sum_a 2T_a^{AA}T_a^{BB}=-\thirt$ (for $A\neq B$). Hence the baryon potential is independent of the color permutation $A,B,C$ and equals
\beq\label{barpot}
V(\yv_{1},\yv_{2},\yv_{3})=\frac{g\la^2}{2\sqrt{6}}\sqrt{(\yv_1-\yv_2)^2+(\yv_2-\yv_3)^2+(\yv_3-\yv_1)^2}
\eeq 
In the limit where two quarks are in the same position, \eg, $\yv_{2} = \yv_{3}$, this potential coincides with the meson potential \eq{linpot}. An analogous potential cannot be constructed for more than three quarks, as there are then more independent quark separations than diagonal SU(3) generators. 

Proceeding as in the $\qu\qb$ case results in the bound state condition on the wave function \eq{colbar} \cite{Hoyer:2009ep},
\beq\label{barbse2}
\sum_{j=1}^3\left[\gamma^0(-i\nabv_{j}\cdot\gv_{j} +m_{j})\right]\chi(\yv_{1},\yv_{2},\yv_{3}) = (E-V)\chi(\yv_{1},\yv_{2},\yv_{3})
\eeq
with $V$ given by \eq{barpot}.

\section{Discussion \label{discuss}}

The above presentation covered much ground. Some of the topics were of a very basic and established nature, while other parts may be classified as recent research proposals. To make it easier to digest the arguments I recapitulate and comment on the main steps.

I discussed a perturbative approach to bound states in field theory. For QED atoms we know that this is the right method, since it gives accurate agreement with data. Applying perturbation theory to bound states requires somewhat different methods than used for scattering amplitude calculations. In order to generate a bound state pole one needs to sum an infinite set of Feynman diagrams. In the $\alpha\to 0$ limit ladder diagrams of the type shown in \fig{fig4} give the leading contribution when the external momenta are scaled with $\alpha$ so as to be compatible with the bound state wave function. The residue of the pole satisfies the Schr\"odinger equation, which defines the Born term in a systematic evaluation of higher order corrections \cite{Lepage:1978hz,Kinoshita:1998jfa}.

 Perturbation theory can give us insights into the properties of relativistic bound states from ``first principles'', and its relevance for QCD bound states merits careful consideration. There are indications that the QCD coupling has a moderate value $\alpha_0 \simeq 0.5$ even at distances of \order{1\ {\rm fm}}. Many issues involving relativistic states have received less attention than they deserve, such as the boost dependence of the  Hydrogen atom wave function in QED \cite{Jarvinen:2004pi}. As I noted in Sec. \ref{dynboost}, this offers an example of how the Lorentz contraction of a classical stick is realized, and differs, in quantum mechanics. 

In perturbation theory already the lowest order contribution is expected to provide a good qualitative description of the exact result. Hence I focussed on the physical principles of bound states at Born level. The Hydrogen atom may be thought of as bound by the instantaneous Coulomb field $A^0(\xv;\xv_1,\xv_2)$ which is determined by Gauss' law \eq{gauss} {\it separately and differently} for each position $\xv_1,\xv_2$ of the electron and proton. Both the electron and the proton interact with this Coulomb field -- double counting is prevented because the field energy \eq{actf} cancels half of the interaction energy. The Coulomb field $A^0$ measured by an external observer far from the bound state is then given by the coherent sum over all configurations of the atom, $A^0(\xv)=\int d\xv_1 d\xv_2 \varphi(\xv_1,\xv_2)A^0(\xv;\xv_1,\xv_2)$ where $\varphi$ is the wave function. 

In courses on relativistic quantum mechanics the Dirac equation is often introduced through its historical context, with a wave function $\varphi(x)$ that is a {\it c-number}. In a field theory context the Dirac equation later reappears as an exact {\it operator-valued} equation derived from the QED action. The similar notation obscures the fact that the $c$-numbered Dirac equation is only an approximate relation which can be derived from summing a subset of Feynman diagrams in a certain limit \cite{Brodsky:1971sk}. The negative energy components of the Dirac wave function $\varphi(x)$ are often vaguely explained as related to antiparticle effects. In Secs. \ref{retprop} and \ref{dirwf} we saw that $\varphi(x)$ is the wave function obtained using retarded propagators and describes a state which, with Feynman propagators, has contributions from an infinite number of particle-antiparticle pairs. The two descriptions give exactly the same bound state energy in the absence of loop corrections. This possibility to describe a state with infinitely many pairs using a single-particle wave function might shed light on the puzzle of why hadron quantum numbers reflect their valence quark degrees of freedom only, with no reference to the sea quarks and gluons.

The use of retarded propagators is only possible in the absence of loops, which in turn is generally understood to imply Born level, or lowest order in $\hbar$. The concept of a Born term for bound states appears not to have been discussed before \cite{Hoyer:2009ep,Brodsky:2010zk}. The Schr\"odinger equation is the Born term for non-relativistic states (Sec. \ref{hbarsect}). The $\hbar$ expansion is equally applicable to relativistic bound states, for which it provides a well-defined first approximation. Since $\hbar$ is a fundamental parameter each order in the expansion must have all the symmetries of the exact result.  As I discussed in Sec. \ref{dynboost}, the full Poincar\'e invariance of quantum theory is not manifest due to the choice of a quantization surface (\eg, equal time or equal light-front time) \cite{Dirac:1949cp}. Hence the Born term of equal-time bound states has a hidden, dynamical boost invariance. This can hardly be obtained without a precise theoretical framework such as perturbation theory \cite{Artru:1983gm}.

Students should be aware that my proposal in Sec.~\ref{confbound} to consider a homogeneous solution to Gauss' law (Eq. \eq{asol2}) is an ``educated speculation''. There is not sufficient experience yet to tell whether such a solution gives fully self-consistent and meaningful results. It fulfills the essential requirement of giving a stationary action, thus maintaining the symmetries of the theory. The bound state equation has manifest rotational symmetry in the CM and dynamic boost invariance for the purely linear potential of the Born term. The application to $\qu\qb$ and $\qu\qu\qu$ bound states in QCD is straightforward and maintains color covariance. Further work is needed to study the properties of such solutions. The present hamiltonian formulation should allow to study many physical observables such as form factors, parton distributions and scattering phenomena. The relevance of imposing the boundary condition \eq{asolqcd} on $A^0(\xv)$ which gives a finite energy density as $|\xv| \to \infty$ will depend on the outcome of those applications.

I should emphasize that the present approach does not explain why there is confinement in QCD but not in QED. The vacuum energy density $\la^4$ in \eq{lamcond} is a free parameter, which must be set to zero ``by hand'' for QED. This is only a proposal for how confinement can be described in a field theory context.

I did not discuss the important issue of chiral symmetry. In the spirit of the present approach a possibility to describe spontaneous symmetry breaking may be to impose chirally non-symmetric states as a boundary condition, in analogy to the non-vanishing of the energy density, $\la \neq 0$ in \eq{lamcond}.

\acknowledgments

I wish to thank the organizers of the International Summer School and Conference on High Energy Physics: Standard Model and Beyond (ISSCSMB '10) for their kind invitation and support. I am also grateful for comments on the present manuscript by Matti J\"arvinen and for travel support from the Magnus Ehrnrooth Foundation.

\end{document}